\documentclass[12pt,twoside]{article}
\usepackage{amssymb}
\usepackage{amsmath}
\usepackage{latexsym}
\usepackage{longtable}
\usepackage{epsfig}
\usepackage{graphicx,bbm,psfrag}
\graphicspath{{images/}}

\begin{document}
\begin{center}
{\Large{\bf Towards a Microscopic Derivation
of the}}\medskip\\
 {\Large{\bf Phonon Boltzmann Equation}}\bigskip\bigskip\\
Herbert Spohn\footnote{{\tt spohn@ma.tum.de}}\medskip\\
Zentrum Mathematik and Physik Department, TU M\"{u}nchen,\\
D - 85747 Garching, Boltzmannstr. 3, Germany\medskip\\
\end{center}
\section{Introduction}\label{sec.1}
The thermal conductivity of insulating (dielectric) crystals is
computed almost exclusively on the basis of the phonon Boltzmann
equation. We refer to \cite{HS} for a discussion more complete
than possible in this contribution. On the microscopic level the
starting point is the Born-Oppenheimer approximation (see
\cite{Teu} for a modern version), which provides an effective
Hamiltonian for the slow motion of the nuclei. Since their
deviation from the equilibrium position is small, one is led to a
wave equation with a \textit{weak} nonlinearity. As already
emphasized by R.~Peierls in his seminal work \cite{Pe}, physically
it is of importance to retain the structure resulting from the
atomic lattice, which forces the discrete wave equation.

On the other hand, continuum wave equations with weak nonlinearity
appear in the description of the waves in the upper ocean and in
many other fields. This topic is referred to as weak turbulence.
Again the theoretical treatment of such equations is based mostly
on the phonon Boltzmann eqation, see e.g.~\cite{Za}. In these
applications one considers scales which are much larger than the
atomistic scale, hence quantum effects are negligible. For
dielectric crystals, on the other side, quantum effects are of
importance at low temperatures. We refer to \cite{HS} and discuss
here only the classical discrete wave equation with a small
nonlinearity.

If one considers crystals with a single nucleus per unit cell,
then the displacement field is a 3-vector field over the crystal
lattice $\Gamma$. The nonlinearity results from the weakly
non-quadratic interaction potentials between the nuclei. As we
will see, the microscopic mechanism responsible for the validity
of the Boltzmann equation can be understood already in case the
displacement field is declared to be scalar, the nonlinearity to
be due to an on-site potential, and the lattice $\Gamma =
\mathbb{Z}^3$. This is the model I will discuss in my notes.

As the title indicates there is no complete proof available for
the validity of the phonon Boltzmann equation. The plan is to
explain the kinetic scaling and to restate our conjecture in terms
of the asymptotics of certain Feynman diagrams.

\section{Microscopic model}\label{sec.2}
\setcounter{equation}{0}

We consider the simple cubic crystal $\mathbb{Z}^3$. The
displacement field is denoted by
\begin{equation}\label{2.1}
q_x \in \mathbb{R}\,, \quad x \in \mathbb{Z}^3\,,
\end{equation}
with the canonically conjugate momenta
\begin{equation}\label{2.2}
p_x \in \mathbb{R}\,, \quad x \in \mathbb{Z}^3\,.
\end{equation}
We use units in which the mass of the nuclei is $m=1$. The
particles interact harmonically and are subject to an on-site
potential, which is divided into a quadratic part and a
non-quadratic correction. Thus the Hamiltonian of the system reads
\begin{equation}\label{2.4}
H= \frac{1}{2} \sum_{x \in \mathbb{Z}^3} \Big( p_x^2 +
\omega_0^2q_x^2\Big)+ \frac{1}{2}\sum_{x,y \in
\mathbb{Z}^3}\alpha(x-y)q_xq_y +\sum_{x \in
\mathbb{Z}^3}V(q_x)=H_0+\sum_{x \in \mathbb{Z}^3}V(q_x)\,.
\end{equation}
The coupling constants have the properties
\begin{equation}\label{2.5}
\alpha(x)=\alpha(-x)\,,
\end{equation}
\begin{equation}\label{2.5a}
|\alpha(x)|\leq\alpha_0 e^{-\gamma|x|}
\end{equation}
for suitable $\alpha_0$, $\gamma>0$, and
\begin{equation}\label{2.5a}
\sum_{x\in \mathbb{Z}^3} \alpha(x)=0\,,
\end{equation}
because of the invariance of the  interaction between the
nuclei under the translation $q_x \leadsto q_x + a$ .

For the anharmonic on-site potential we set
\begin{equation}\label{2.6}
V(u)=\sqrt{\varepsilon}\frac{1}{3}\lambda u^3
+\varepsilon(\lambda^2/18 \omega^2_0)u^4\,,\; u\in \mathbb{R}\,.
\end{equation}
$\varepsilon$ is the dimensionless scale parameter, eventually
$\varepsilon\to 0$. The quartic piece is added so to make sure
that $H\geq 0$. In the limit $\varepsilon\to 0$ its contribution
will vanish and for simplicity of notation we will omit it from
the outset. Then the equations of motion are
\begin{eqnarray}\label{2.7}
&&\frac{d}{dt}q_x (t) = p_x (t)\,,\nonumber\\
&&\frac{d}{dt}p_x (t) = -\sum_{y\in \mathbb{Z}^3}\alpha(y-x)q_y(t)
- \omega_0^2 q_x (t)- \sqrt{\varepsilon}\lambda q_x (t)^2\,, \quad
x \in \mathbb{Z}^3\,.
\end{eqnarray}
We will consider only finite energy solutions. In particular, it
is assumed that $|p_x|\to 0$, $|q_x|\to 0$ sufficiently fast as
$|x|\to \infty$. In fact, later on there will be the need to
impose random initial data, which again are assumed to be
supported on finite energy configurations. In the kinetic limit
the average energy will diverge as $\varepsilon^{-3}$.

It is convenient to work in Fourier space. For $f:\mathbb{Z}^3 \to
\mathbb{R}$ we define
\begin{equation}\label{2.8}
\widehat{f}(k)=\sum_{x\in\mathbb{Z}^3}e^{-i2\pi k\cdot
x}f_x\,,
\end{equation}
$ k\in \mathbb{T}^3 = [ - \frac{1}{2},\frac{1}{2}]^3$, with inverse
\begin{equation}\label{2.9}
f_x=\int_{\mathbb{T}^3}dk e^{ i 2\pi k\cdot x}\widehat{f}(k)\,,
\end{equation}
$dk$ the 3-dimensional Lebesgue measure. The dispersion relation
for the harmonic part $H_0$ is
\begin{equation}\label{2.8}
\omega(k) = \big(\omega_0^2 + \widehat{\alpha}(k)\big)^{1/2}\geq
\omega_0>0\,,
\end{equation}
since $\widehat{\alpha}(k)>0$ for  $k\neq 0$ because of the
mechanical stability of the harmonic lattice with vanishing
on-site potential.

In Fourier space the equations of motion read
\begin{eqnarray}\label{2.10}
&&\hspace{-26pt}\frac{\partial}{\partial t}\widehat{q}(k,t) = \widehat{p}(k,t)\,,\nonumber\\
&&\hspace{-26pt}\frac{\partial}{\partial t}\widehat{p}(k,t) = -
\omega (k)^2
\widehat{q}(k,t)\nonumber\\
&&\hspace{36pt}-\sqrt{\varepsilon}\lambda \int_{\mathbb{T}^6}dk_1
dk_2 \delta(k-k_1-k_2) \widehat{q}(k_1,t) \widehat{q}(k_2,t)
\end{eqnarray}
with $k\in \mathbb{T}^3$. Here $\delta$ is the $\delta$-function
on the unit torus, to say, $\delta (k')$ carries a point mass
whenever $k'\in\mathbb{Z}^3$.

It will be convenient to concatenate $q_x$ and $p_x$ into a single
complex-valued field. We set
\begin{equation}\label{2.11}
a(k)= \frac{1}{\sqrt{2}} \big(\sqrt{\omega} \widehat{q}(k) + i
\frac{1}{\sqrt{\omega}} \widehat{p}(k)\big)
\end{equation}
with the inverse
\begin{equation}\label{2.12}
\widehat{q}(k) = \frac{1}{\sqrt{2}}\frac{1}{\sqrt{\omega}}\big(
a(k) + a(-k)^\ast\big)\,,\; \widehat{p}(k) = \frac{1}{\sqrt{2}} i
{\sqrt{\omega}}\big(-a(k) + a(-k)^\ast \big)\,.
\end{equation}
To have a concise notation, we introduce
\begin{equation}\label{2.13}
a(k,+)= a(k)^\ast \,,\;a(k,-)=a(k)\,.
\end{equation}
Then the $a$-field evolves as
\begin{eqnarray}\label{2.14}
&&\hspace{-10pt}\frac{\partial}{\partial t} a(k,\sigma,t) = i
\sigma \omega(k) a(k,\sigma,t)+ i \sigma \sqrt{\varepsilon}\lambda
\sum_{\sigma_1,\sigma_2=\pm1} \int_{\mathbb{T}^6}d k_1
d k_2\nonumber\\
&&\hspace{5pt}(8 \omega(k) \omega(k_1) \omega(k_2))^{-1/2}
\delta(-\sigma k+\sigma_1k_1+\sigma_2k_2) a(k_1,\sigma_1,t)
a(k_2,\sigma_2,t)\,.
\end{eqnarray}

\section{Kinetic limit and Boltzmann equation}\label{sec.3}
\setcounter{equation}{0}

The kinetic limit deals with a special class of initial
probability measures. Their displacement field has a support of
linear size $\varepsilon^{-1}$ and average energy of order
$\varepsilon^{-3}$. More specifically, these probability measures
have the property of being locally Gaussian and almost stationary
under the dynamics. Because of the assumed slow variation in space
the covariance of such probability measures changes only slowly,
i.e.~on the scale $\varepsilon^{-1}$, in time.

Let us assume then that the initial data for (\ref{2.14}) are random and
specified by a
Gaussian probability measure on phase space. It is assumed to have mean
\begin{equation}\label{3.1}
\langle a(k,\sigma)\rangle_\varepsilon^\textrm{G} =0\,
\end{equation}
and for the covariance we set
\begin{equation}\label{3.2}
\langle a(k,\sigma) a(k',\sigma) \rangle_\varepsilon^\textrm{G}
=0\,,
\end{equation}
\begin{equation}\label{3.3}
W^{\varepsilon} (y,k) = \varepsilon^3
\int_{(\mathbb{T}/\varepsilon)^3} d\eta e^{ i 2\pi
y\cdot\eta}\langle a(k-\varepsilon \eta/2,+) a(k+ \varepsilon
\eta/2,-)\rangle_\varepsilon^\textrm{G}\,,
\end{equation}
$y\in(\varepsilon\mathbb{Z})^3$, which defines the \textit{Wigner
function} rescaled to the lattice $(\varepsilon\mathbb{Z})^3$.
Local stationarity is ensured by the condition
\begin{equation}\label{3.4}
\lim_{\varepsilon \to 0} W^{\varepsilon}(\lfloor
r\rfloor_\varepsilon,k) = W^0(r,k)\,,
\end{equation}
where $\lfloor r\rfloor_\varepsilon$ denotes integer part modulo
$\varepsilon$. Note that $W^\varepsilon$ is normalized as
\begin{equation}\label{3.5}
\sum_{y\in (\varepsilon \mathbb{Z})^3}\int_{\mathbb{T}^3}dk
W^{\varepsilon}(y,k) = \int_{\mathbb{T}^3}dk \langle a(k,+) a(k,-)
\rangle^\textrm{G}_\varepsilon\,.
\end{equation}
The condition that the limit in (\ref{3.4}) exists thus implies
that the average phonon number increases as $\varepsilon^{-3}$,
equivalently the average total energy increases as
\begin{equation}\label{3.18}
\langle\int_{\mathbb{T}^3}d^3k \omega(k) a(k,+) a(k,-)
\rangle_\varepsilon^\textrm{G} = \langle
H_0\rangle_\epsilon^\textrm{G}= \mathcal{O}(\varepsilon^{-3})\,.
\end{equation}

Let $\langle \cdot\rangle_t$ be the time-evolved measure at time
$t$. Its rescaled Wigner function is
\begin{eqnarray}\label{3.6}
W^{\varepsilon} (y,k,t) = \varepsilon^3
\int_{(\mathbb{T}/\varepsilon)^3} d\eta e^{ i 2\pi
y\cdot\eta}\langle a(k- \varepsilon \eta/2,+) a(k+ \varepsilon
\eta/2,-)\rangle_{t/\varepsilon}\,.
\end{eqnarray}
Kinetic theory claims that
\begin{equation}\label{3.7}
\lim_{\varepsilon\to 0}W^\varepsilon (\lfloor r
\rfloor_\varepsilon, k,t) = W(r,k,t)\,,
\end{equation}
where $W(r,k,t)$ is the solution of the phonon Boltzmann equation
\begin{eqnarray}\label{3.8}
&&\hspace{-20pt}\frac{\partial}{\partial t} W(r,k,t) + \frac{1}{2\pi} \nabla
\omega (k)\cdot \nabla_r W( r,k,t)\nonumber\\
&&\hspace{-5pt}= \frac{\pi}{2}\lambda^2 \sum_{\sigma_1,\sigma_2=\pm 1}
\int_{\mathbb{T}^6} d k_1 d k_2 (\omega(k) \omega(k_1)
\omega(k_2))^{-1} \delta
(\omega(k)+\sigma_1\omega(k_1)+\sigma_2\omega(k_2))\nonumber\\
&&\hspace{80pt}\delta(k+\sigma_1 k_1+\sigma_2k_2)\big(W(r,k_1,t)
W(r,k_2,t)\nonumber\\
&&\hspace{80pt} + \sigma_1
W(r,k,t)W(r,k_2,t)+\sigma_2W(r,k,t)W(r,k_1,t)\big)
\end{eqnarray}
to be solved with the initial condition $W(r,k,0)=W^0(r,k)$.

The free streaming part is an immediate consequence of the
evolution of $W$ as generated by $H_0$. The strength of the cubic
nonlinearity was assumed to be of order $\sqrt{\varepsilon}$,
which results in an effect of order 1 on the kinetic time scale.
The specific form of the collision operator will be explained in
the following section. It can be brought into a more familiar form
by performing the sum over $\sigma_1,\sigma_2$. Then the collision
operator has two terms. The first one describes the merging of two
phonons with wave number $k$ and $k_1$ into a phonon with wave
number $k_2=k+k_1$, while the second term describes the splitting
of a phonon with wave number $k$ into two phonons with wave
numbers $k_1$ and $k_2$, $k=k_1+k_2$. In such a collision process
energy is conserved and wave number is conserved modulo an integer
vector.

In (\ref{4.7}) the summand with $\sigma_1 =1= \sigma_2$ vanishes
trivially. However it could be the case that the condition for
energy conservation,
\begin{equation}\label{3.9}
\omega(k) + \omega(k') - \omega(k+k')=0\,,
\end{equation}
has also no solution. If so, the collision operator vanishes. In
fact, for nearest neighbor coupling only, $\alpha(0)=6$,
$\alpha(e)=-1$ for $|e|=1$, $\alpha(x)=0$ otherwise, it can be
shown that (\ref{3.9}) has no solution whenever $\omega_0 > 0$. To
have a non-zero collision term we have to require
\begin{equation}\label{3.10}
\int dk \int dk' \delta(\omega(k)+\omega(k')-\omega(k+k'))>0\,,
\end{equation}
which is an implicit condition on the couplings $\alpha(x)$. A
general condition to ensure (\ref{3.10}) is not known. A simple
example where (\ref{3.10}) can be checked by hand is
\begin{equation}\label{3.11}
\omega(k)=\omega_0 + \sum^3_{\alpha=1} (1-\cos(2\pi
k^\alpha))\,,\quad k=(k^1,k^2,k^3)\,.
\end{equation}
It corresponds to suitable nearest and next nearest neighbor
couplings.

There is a second more technical condition which requires that
\begin{equation}\label{3.12}
\sup_k \int dk'\delta
(\omega(k)+\omega(k')-\omega(k+k'))=c_0<\infty\,.
\end{equation}
It holds for the dispersion relation (\ref{3.11}). This uniform
bound allows for a simple proof that the Boltzmann equation has a
unique solution for short times provided $W^0(r,k)$ is bounded.

\section{Feynman diagrams}\label{sec.4}
\setcounter{equation}{0}

Denoting by $\langle\cdot\rangle_t$ the average with respect to
the measure at time $t$ (in microscopic units), the starting point
of the time-dependent perturbation series is the identity
\begin{eqnarray}\label{4.1}
&&\hspace{-60pt} \langle\prod^n_{j=1} a(k_j,\sigma_j)\rangle_t =
\exp\big[it\big(\sum^n_{j=1} \sigma_j\omega(k_j)\big)\big]\langle
\prod^n_{j=1}
a(k_j,\sigma_j)\rangle^\textrm{G}\nonumber\\
&&\hspace{26pt} + i\sqrt{\varepsilon}\int^t_0 ds
\exp\big[i(t-s)\big(\sum^n_{j=1} \sigma_j\omega(k_j)\big)\big]\nonumber\\
&&\hspace{26pt}\Big(\sum^n_{\ell=1}
\sum_{\sigma',\sigma''=\pm1}\sigma_\ell
\int_{\mathbb{T}^6} dk'dk''\phi(k_\ell,k',k'')\delta(-\sigma_\ell k_\ell+\sigma'k'+\sigma''k'')\nonumber\\
&&\hspace{26pt} \langle (\prod^n_{j=1\;j\neq\ell} a(k_j,\sigma_j))
a(k',\sigma')a(k'',\sigma'')\rangle_s\Big)\,.
\end{eqnarray}
Here
\begin{equation}\label{4.2}
\phi(k,k',k'')=\lambda(8\omega(k)\omega(k')\omega(k''))^{-1/2}
\end{equation}
One starts with $n=2$ and $\sigma_1=1$, $\sigma_2=-1$. Then on the
right hand side of (\ref{4.1}) there is the product of three
$a$'s. One resubstitutes (\ref{4.1}) with $n=3$, etc. Thereby one
generates an infinite series, in which only the average over the
initial Gaussian measure $\langle\cdot\rangle^\textrm{G}$ appears.

To keep the presentation transparent, let me assume that
$\langle\cdot\rangle^\textrm{G}$ is a translation invariant
Gaussian measure with
\begin{eqnarray}\label{4.3}
&&\hspace{-30pt}\langle a(k,\pm)\rangle^\textrm{G}=0\,,\;\langle
a(k,\sigma)a(k',\sigma)\rangle^\textrm{G}=0\,,\nonumber\\
&&\hspace{-30pt}\langle a(k,+)a(k',-)\rangle^\textrm{G}
=\delta(k-k') W(k)\,.
\end{eqnarray}
Then the measure at time $t$ is again translation invariant.
Kinetic scaling now merely amounts to considering the long times
$t/\varepsilon$. The Wigner function at that time is then
represented by the infinite series
\begin{equation}\label{4.4}
\langle a(q,-)a(p,+)\rangle_{t/\varepsilon}=\delta(q-p)
\Big(W(q)+\sum^\infty_{n=1}W^\varepsilon_n(q,t)\Big)\,.
\end{equation}
The infinite sum is only formal. Taking naively the absolute value
at iteration $2n$ one finds that
\begin{equation}\label{4.5}
|W^\varepsilon_n(q,t)|\leq
\varepsilon^n(t/\varepsilon)^{2n}((2n)!)^{-1}(2n)!
c^{2n}((2n+2)!/2^{n+1}(n+1)!)\,.
\end{equation}
Here $\varepsilon^n=(\sqrt{\varepsilon})^{2n}$,
$(t/\varepsilon)^{2n}/(2n)!$ comes from the time integration,
$(2n)!$ from the sum over $\ell$ in (\ref{4.1}), $c^{2n}$ from the
$k$-integrations and the initial $W(k)$, and the factor
$(2n+2)!/2^{n+1}(n+1)!$ from the Gaussian pairings in the
initial measure. Thus even at fixed $\varepsilon$ there are too
many terms in the sum.

Since no better estimate is available at present, we concentrate
on the structure of a single summand $W^\varepsilon_n(q,t)$.
$\delta(q-p)W^\varepsilon_n(q,t)$ is a sum of integrals. The
summation comes from\smallskip\\
-- the sum over $\sigma',\sigma''$ in (\ref{4.1})\\
-- the sum over $\ell$ in (\ref{4.1})\\
-- the sum over all pairings resulting from the average with
respect to the initial Gaussian measure
  $\langle\cdot\rangle^\textrm{G}$.\smallskip\\
Since each single integral has a rather complicated structure, it
is convenient to visualize them as \textit{Feynman diagrams}.

A Feynman diagram is a graph with labels. Let us first explain the
graph. The graph consists of two binary trees. It is convenient to
draw them on a ``backbone" consisting of $2n+2$ equidistant
horizontal level lines which are labelled from 0 (bottom) to
$2n+1$ (top). The two roots of the tree are two vertical bonds
from line $2n+1$ to level line $2n$. At level $m$ there is
\textit{exactly one} branch point with two branches in either
tree. Thus there are exactly $2n$ branch points. At level 0 there
are then $2n+2$ branches. They are connected according to the
pairing rule, see Figure below.

In the Feynman graph each bond is oriented with arrows pointing
either up $(\sigma=+1)$ or down $(\sigma=-1)$. The left root is
down while the right root is up. If there is no branching the
orientation is inherited from the upper level. At a pairing the
orientation must be maintained. Thus at level 0 a branch with an
up arrow can be paired only with a branch with a down arrow, see
(\ref{4.3}). Every internal line in the graph must terminate at
either end by a branch point. Every such internal line admits
precisely two orientations.

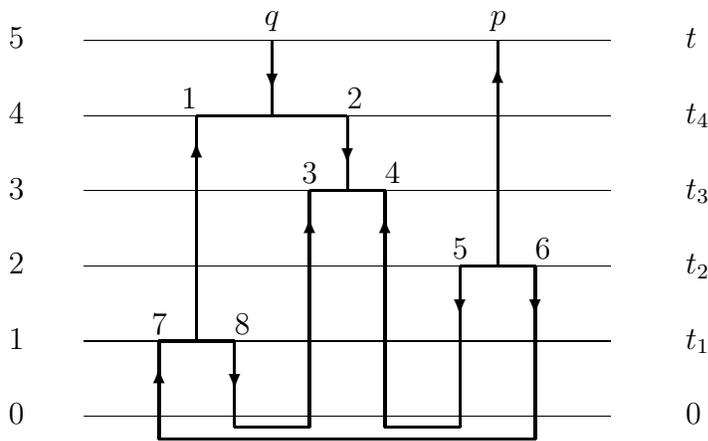
\setlength{\unitlength}{1cm}
\begin{picture}
(10,8)(-3.5,0)
\put(-1,-0.1){$0$} \put(-1,0.9){$1$} \put(-1,1.9){$2$}
\put(-1,2.9){$3$} \put(-1,3.9){$4$} \put(-1,4.9){$5$}
\put(8,-0.1){$0$} \put(8,0.9){$t_1$} \put(8,1.9){$t_2$}
\put(8,2.9){$t_3$} \put(8,3.9){$t_4$} \put(8,4.9){$t$}

\put(1.3,4.1){$1$}\put(3.5,4.1){$2$}\put(2.9,3.1){$3$}\put(4,3.1){$4$}
\put(4.9,2.1){$5$}\put(6,2.1){$6$}\put(0.9,1.1){$7$}\put(2,1.1){$8$}
\put(2.4,5.2){$q$}\put(5.4,5.2){$p$}
\linethickness{0.1pt}
\put(0,0){\line(1,0){7}}
\put(0,1){\line(1,0){7}}
\put(0,2){\line(1,0){7}}
\put(0,3){\line(1,0){7}}
\put(0,4){\line(1,0){7}}
\put(0,5){\line(1,0){7}}
\thicklines
\put(1,-0.3){\line(0,1){1.3}}
\put(2,-0.15){\line(0,1){1.15}}
\put(3,-0.15){\line(0,1){3.15}}
\put(4,-0.15){\line(0,1){3.15}}
\put(5,-0.15){\line(0,1){2.15}}
\put(6,-0.3){\line(0,1){2.3}}
\put(1.5,1){\line(0,1){3}}
\put(2.5,4){\line(0,1){1}}
\put(3.5,3){\line(0,1){1}}
\put(5.5,2){\line(0,1){3}}
\put(1,1){\line(1,0){1}}
\put(3,3){\line(1,0){1}}
\put(1.5,4){\line(1,0){2}}
\put(5,2){\line(1,0){1}}
\put(1,-0.3){\line(1,0){5}}
\put(2,-0.15){\line(1,0){1}}
\put(4,-0.15){\line(1,0){1}}
\put(1.5,3.5){\vector(0,1){0.15}}
\put(1,0.5){\vector(0,1){0.15}}
\put(2,0.5){\vector(0,-1){0.15}}
\put(2.5,4.5){\vector(0,-1){0.15}}
\put(3,2.5){\vector(0,1){0.15}}
\put(4,2.5){\vector(0,1){0.15}}
\put(5,1.5){\vector(0,-1){0.15}}
\put(6,1.5){\vector(0,-1){0.15}}
\put(5.5,4.5){\vector(0,1){0.15}}
\put(3.5,3.5){\vector(0,-1){0.15}}

\end{picture}\\

\newpage

Next we insert the labels. The level lines 0 to $2n+1$ are
labelled by times $0<t_1\ldots<t_{2n}<t$. The left root carries
the label $q$ while the right root carries the label $p$. Each
internal line is labelled with a wave number $k$.

To each Feynman diagram one associates an integral through the
following steps.\smallskip\\
(i) The time integration is over the simplex $0\leq
t_1\ldots\leq t_{2n}\leq t$ as $dt_1\ldots dt_{2n}$.\smallskip\\
(ii) The wave number integration is over all internal lines as
$\int dk_1 \ldots \int dk_\kappa$, where $\kappa=3n-1$ is the
number of internal
lines.\smallskip\\
(iii) One sums over all orientations of the internal lines.\smallskip\\
The integrand is a product of three factors.\smallskip\\
(iv) There is a product over all branch points.
At each branchpoint there is a root, say wave vector $k_1$ and
orientation $\sigma_1$, and there are two branches, say wave vectors $k_2$,
$k_3$ and orientations $\sigma_2$, $\sigma_3$. Then each branch
point carries the weight
\begin{equation}\label{4.6}
\delta(-\sigma_1k_1+\sigma_2k_2+\sigma_3k_3)\sigma_1\phi(k_1,k_2,k_3)\,.
\end{equation}
If one regards the wave vector $k$ as a current with orientation
$\sigma$, then (\ref{4.6}) expresses Kirchhoff's rule for
conservation of the current.\smallskip\\
(v) By construction each bond carries a time difference
$t_{m+1}-t_m$, a wave vector $k$, and an orientation $\sigma$.
Then to this bond one associates the phase factor
\begin{equation}\label{4.7}
\exp[i(t_{m+1}-t_m)\sigma\omega(k)/\varepsilon]\,.
\end{equation}
The second factor is the product of such phase factors over all bonds.\smallskip\\
(vi) The third factor of the integrand is given by
\begin{equation}\label{4.8}
\prod^{n+1}_{j=1} W(k_j)\,,
\end{equation}
where $k_1,\ldots,k_{n+1}$ are the wave numbers of the bonds
between level 0 and level 1.\smallskip\\
(vii) Finally there is the prefactor
$(-1)^n\varepsilon^{-n}$.\smallskip

To illustrate these rules we give an example for $n=2$, see Figure
above. The associated integral is given by,  more transparently keeping the $\delta$-functions
from the pairings,
\begin{eqnarray}\label{4.8a}
&&\hspace{-8pt}\varepsilon^{-2}\int_{0\leq t_1\leq\ldots\leq t_4 \leq
t}dt_1\ldots dt_4
\int_{\mathbb{T}^{24}}dk_1\ldots dk_8\nonumber\\
&&\delta(q+k_1-k_2)\delta(k_2+k_3+k_4)\delta(-p-k_5-k_6)
\delta(-k_1+k_7-k_8)\nonumber\\
&&\phi(q,k_1,k_2)\phi(k_2,k_3,k_4)\phi(p,k_5,k_6)\phi(k_1,k_7,k_8)\nonumber\\
&&\hspace{0pt}\delta(k_7-k_6)W(k_7)\delta(k_8-k_3)W(k_8)\delta(k_4-k_5)W(k_4)\nonumber\\
&&\exp\big[\{i(t-t_4)(-\omega_q+\omega_p)+
i(t_4-t_3)(\omega_1-\omega_2+\omega_p)\nonumber\\
&&+i(t_3-t_2)(\omega_1+\omega_3+\omega_4+\omega_p)
+i(t_2-t_1)(\omega_1+\omega_3+\omega_4-\omega_5-\omega_6)\nonumber\\
&& +it_1(\omega_7-\omega_8+\omega_3+\omega_4-\omega_5-\omega_6)\}/\varepsilon\big]
\end{eqnarray}
with $\omega_q=\omega(q)$, $\omega_p=\omega(p)$,
$\omega_j=\omega(k_j)$.

$\delta(q-p)W^\varepsilon_n(q,t)$ is the sum over all Feynman
diagrams with $2n+2$ levels and thus is a sum of oscillatory
integrals. In the limit $\varepsilon\to 0$ only a few leading
terms survive while all remainders vanish. E.g., the Feynman
diagram above is subleading. In fact, the conjecture
of kinetic theory can be stated rather concisely:\bigskip\\
\textbf{Kinetic Conjecture}: \textit{
In a leading Feynman diagram
the Kirchhoff rule never forces an identification of the form $\delta(k_j)$ with some wave vector $k_j$. In addition, 
the sum of the $2(n-m+1)$ phases from the bonds between level lines $2m$
and $2m+1$ vanishes for every choice of internal wave
numbers. This cancellation must hold for $m=0,\ldots,n$.}\bigskip

Since we assumed the initial data to be spatially homogeneous, the
phonon Boltzmann equation (\ref{3.8}) simplifies to
\begin{eqnarray}\label{4.9}
&&\hspace{-16pt}\frac{\partial}{\partial t} W(k,t)= 4\pi \lambda^2
\sum_{\sigma_1,\sigma_2=\pm1} \int_{\mathbb{T}^6} d k_1 d k_2
\phi(k,k_1,k_2)^2 \delta
(\omega(k)+\sigma_1\omega(k_1)+\sigma_2\omega(k_2))\nonumber\\
&&\hspace{16pt}\delta(k+\sigma_1 k_1+\sigma_2k_2)\big(W(k_1,t)
W(k_2,t) + 2\sigma_2 W(k,t)W(k_1,t)\big)\,,
\end{eqnarray}
where we used the symmetry with respect to $(k_1,\sigma_1)$ and
$(k_2,\sigma_2)$. To (\ref{4.9}) we associate the Boltzmann
hierarchy
\begin{equation}\label{4.9a}
\frac{\partial}{\partial t}f_n= \mathcal{C}_{n,n+1} f_{n+1}\,,\;
n=1,2,\ldots\,,
\end{equation}
acting on the symmetric functions $f_n(k_1,\ldots,k_n)$ with
\begin{eqnarray}\label{4.10}
&&\hspace{-16pt}\mathcal{C}_{n,n+1}f_{n+1}(k_1,\ldots,k_n)=4\pi\lambda^2\sum^n_{\ell=1}
\sum_{\sigma',\sigma''=\pm 1}\int_{\mathbb{T}^6} d k' d k''
\phi(k_\ell,k',k'')^2\nonumber\\
&&\hspace{16pt}
\delta(\omega(k_\ell)+\sigma'\omega(k')+\sigma''\omega(k''))
\delta(k_\ell+\sigma'k'+\sigma''k'')\nonumber\\
&&\hspace{16pt}
[f_{n+1}(k_1,\ldots,k',k_{\ell+1},\ldots,k'')+2\sigma''
f(k_1,\ldots,k_n,k')]\,.
\end{eqnarray}
Under the condition (\ref{3.12}) and provided
$\|W\|_\infty<\infty$, the hierarchy (\ref{4.9a}) has a unique
solution for short times. In case
\begin{equation}\label{4.11}
f_n(k_1,\ldots,k_n,0)=\prod^n_{j=1}W(k_j)\,,
\end{equation}
the factorizatin is maintained in time and each factor agrees with
the solution of the Boltzmann equation (\ref{4.9}). From
(\ref{4.9a}) one easily constructs the perturbative solution to
(\ref{4.9}) with the result
\begin{eqnarray}\label{4.12}
&&\hspace{-16pt}W(k,t)=W(k)+\sum^\infty_{n=1}
\frac{1}{n!}t^n(\mathcal{C}_{1,2}\ldots
\mathcal{C}_{n,n+1}W^{\otimes n+1})(k)\nonumber\\
&&\hspace{26pt} =W(k)+\sum^\infty_{n=1}W_n(k,t)\,.
\end{eqnarray}
The series in (\ref{4.12}) converges for $t$ sufficiently small.

For $n=1,2$ the oscillating integrals can be handled by direct
inspection with the expected results $\lim_{\varepsilon\to
0}W^\varepsilon_1(k,t)=W_1(k,t)$, $\lim_{\varepsilon\to
0}W^\varepsilon_2(k,t)=W_2(k,t)$. If the leading terms are as
claimed in the Kinetic Conjecture, then they agree with the series
(\ref{4.12}). The complete argument is a somewhat tricky counting
of diagrams, which would lead us too far astray. Thus the most
immediate project is to establish that all subleading diagrams
vanish in the limit $\varepsilon\to 0$. This would be a step
further when compared to the investigations \cite{BeCa},
\cite{ErSaY}.

Of course a complete proof must deal with the uniform convergence
of the series in (\ref{4.4}).\bigskip\\
{\bf Acknowledgements}. I thank Jani Lukkarinen for instructive
discussions and Gianluca Panati for a careful reading.

\end{document}